\documentclass[preprint,12pt]{elsarticle}

\usepackage{amssymb}
\usepackage{amsmath}
\usepackage{booktabs}
\usepackage{multicol}
\usepackage{multirow}
\usepackage{bm}

\journal{Displays}

\begin{document}

\begin{frontmatter}



\title{Visual-Motion-Induced Modulation of Pedestrian Trajectories Using Spatially Distributed Multi-Display Signage in Public Spaces}


\author[label1]{Yuri Mikawa}
\author[label2]{Taiki Fukiage} 
\author[label2]{Yuki Kubota}
\author[label2]{Takumi Yokosaka}
\author[label2]{Maki Ogawa}
\author[label2]{Kazushi Maruya}

\affiliation[label1]{organization={The University of Tokyo},
            addressline={7-3-1, Hongo}, 
            city={Bunkyo-ku},
            postcode={113-8656}, 
            state={Tokyo},
            country={Japan}}
\affiliation[label2]{organization={NTT, Inc.},
            addressline={3-1, Morinosato-Wakamiya}, 
            city={Atsugi-shi},
            postcode={243-0198}, 
            state={Kanagawa},
            country={Japan}}

\begin{abstract}

Multi-display signage (MDS), now ubiquitous in urban environments, has the potential to influence human behavior and experience in public spaces. However, despite its unique capability to present spatially distributed dynamic visual stimuli, its current use is mainly limited to advertising. 

In this study, we propose a perception-based approach for laterally modulating pedestrian trajectories as a nonverbal means of guiding pedestrians in public spaces. 
The approach is motivated by vection—the illusion of self-motion—and uses laterally moving monochrome stripes, a standard stimulus in vection research, presented across spatially distributed displays to elicit postural responses that may bias pedestrian trajectories.

We evaluated the approach through a controlled laboratory experiment and a real-world field deployment involving actual pedestrian flows in a national museum. The laboratory experiment examined whether the MDS setup induced trajectory shifts in the direction predicted by prior research on the behavioral effects of vection. The field deployment investigated whether comparable effects would emerge in aggregate pedestrian behavior during unconstrained movement under conditions closer to those of urban public spaces.

In the laboratory, full-screen motion significantly biased walking trajectories in the direction of visual motion, whereas partial-stripe motion produced no significant directional effect. In the field deployment, opposing full-screen motion conditions produced direction-consistent differences in aggregate pedestrian positions. 
The field results, observed despite the substantial variability in real-world pedestrian flows, extend the controlled laboratory findings and provide ecologically valid evidence supporting practical MDS-based pedestrian modulation in public settings.
The results further suggest that sufficient visual-motion coverage may be important.

\end{abstract}




\end{frontmatter}




\begin{figure}[t]
\centering
  \includegraphics[width=\linewidth]{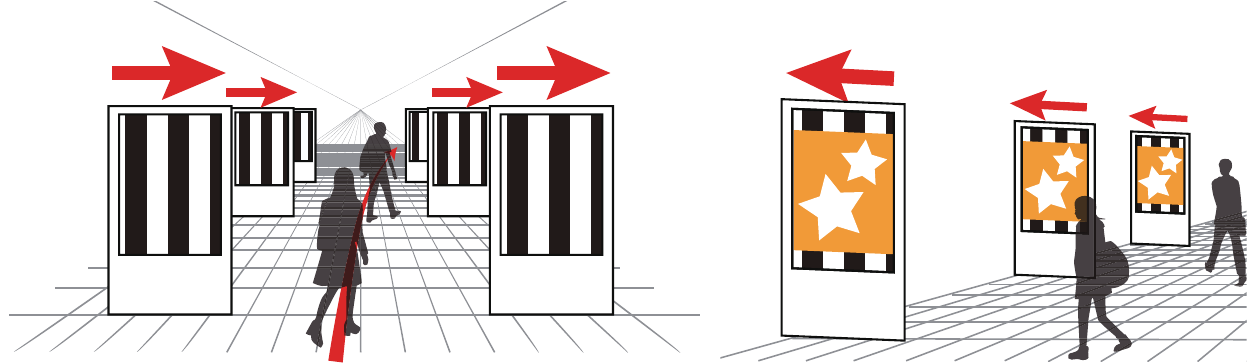}
    \caption{Overview of the visual stimuli used in the study. Vection-inducing visual stimuli are presented on multi-display signage (MDS) to modulate pedestrian trajectories. Left: a full-screen stimulus covering the entire display, as commonly used in vection research. Right: an advertisement-compatible design with narrow moving stripes embedded at the top and bottom of the display while preserving the central advertisement.}
  \label{fig:teaser}
\end{figure}
\section{Introduction}

In contemporary urban public spaces such as train stations, airports, and plazas, multi-display signage (MDS) has become ubiquitous for presenting impactful advertisements to pedestrians \cite{mirgorodskaya2020using}. MDS can be interpreted as a large-field, spatially distributed, and dynamic visual stimulus. These characteristics are largely unique among visual elements in public spaces, as traditional architectural surfaces cannot provide such rich and dynamic visual information \cite{mikawa-APMAR2025}. 
Although MDS is currently used primarily for the repetitive presentation of advertisements, it is also a unique display medium in public spaces in that it can coexist with the real environment while presenting distributed spatial digital imagery to multiple viewers simultaneously without wearable devices. In this sense, MDS can be regarded as a shared spatial display system with the potential to shape the behaviors and experiences of numerous pedestrians in urban public spaces.

One potential application of this capability is pedestrian guidance.
In modern urban spaces, pedestrian movement is commonly managed through explicit signs, arrows, and floor markings~\cite{kataoka2016dynamic}. However, such signage requires active attention and rapid interpretation, and may therefore be overlooked or misunderstood, reducing its intended guidance effect \cite{mishler2017improving,galea2017international}.
Researchers have also explored nonverbal approaches to influence pedestrian movement. For example, some studies have attempted to regulate natural pedestrian flow using floor-mounted visual devices that induce vection---a visually induced perception of self-motion \cite{furukawa2011vection,shimizu2024master,shimizu2024gentlepoles}. 
Although such approaches are promising, they require the installation of additional devices on the floor. Given the widespread deployment of MDS in public spaces, and their ability to cover the peripheral visual field in which such visual motion cues are likely to be effective \cite{brandt1973differential}, reusing existing MDS infrastructure may offer a practical means of delivering perceptual walking cues without additional floor-mounted devices.

In this study, we investigate a method for modulating pedestrian trajectories using spatially distributed visual stimuli presented on MDS in urban public spaces, as shown in Figure~\ref{fig:teaser}. Our approach aims to guide pedestrian behavior by presenting visual-motion stimuli designed to evoke vection. Specifically, we use laterally moving monochrome stripe patterns, which are commonly employed in basic vection research \cite{sauvan1993properties,sauvan1995spatiotemporal,riecke2011compelling,wang2021visually}. Based on prior findings, we hypothesized that such stimuli could produce visually induced postural responses (VIPRs), including postural drift \cite{hanssens2013visually}, and thereby bias pedestrian trajectories through perceptual mechanisms. We examined two types of visual stimuli: (1) full-screen moving stripes presented across the entire display, and (2) partial moving stripes shown only in the upper and lower regions of the display so that they remain compatible with existing advertisements. 
To meet the operational and ethical requirements of this initial study involving members of the public, we temporarily installed an MDS at a national science museum with the museum’s support.

Both a controlled laboratory experiment and a public field trial were conducted, each serving a distinct role in evaluating the proposed approach. The laboratory experiment first examined whether the MDS setup could produce the effects anticipated based on prior research under controlled conditions. Using the laboratory findings as a reference, the field trial then investigated whether similar effects would emerge during unconstrained walking in a real-world environment. The field trial was conducted in a free-admission area of a national museum, where pedestrians had no prescribed starting positions or destinations and social factors, such as group walking, were naturally present. 

The results showed that full-screen motion produced significant directional effects not only under controlled laboratory conditions but also during unconstrained walking in a real-world environment, supporting the ecological relevance of the proposed approach and highlighting the potential of spatially distributed visual stimuli delivered through MDS to modulate pedestrian behavior in public spaces. Evidence from field studies in public spaces is inherently scarce because opportunities to conduct such experiments are limited, making the present findings particularly valuable.
Furthermore, no significant directional effect was detected under the partial-stripe condition in the laboratory experiment, suggesting that sufficient visual-motion coverage, e.g., by integrating motion into the advertisement itself, may be important in practical MDS application.

\section{Related Works}
\subsection{Vection and Visually Induced Postural Responses}


Vection refers to the illusion of self-motion that occurs when a large, coherent visual motion pattern is presented across the retina \cite{palmisano2015future}. A classic example is that people feel their own train is moving in the opposite direction when a neighboring train begins to move. 
The process of self-motion perception consists of several stages. Optic flow is first processed along the visual motion pathways at MT (middle temporal area) and MST (medial superior temporal area) \cite{andersen2000neural,kovacs2008neural}. The visual signals are then integrated with vestibular and somatosensory inputs to estimate the direction of actual self-motion \cite{dichgans1978visual,deangelis2012visual,chen2011convergence}. When visual motion signals conflict with vestibular and somatosensory inputs, the observer may attribute the visual motion to self-motion. 
In particular, vection is known to be strengthened when the optic flow is congruent with actual head motion or the direction of gravity, and weakened when it is incongruent \cite{ash2011vection,palmisano2015future,palmisano2018search}.

Many studies on vection have used monochrome stripe patterns that move horizontally or vertically \cite{furukawa2011vection,sakamoto2019guided}, or random dot patterns that move in depth \cite{wang2021visually,palmisano2004jitter}. The vection effects and optimal conditions to induce vection have been extensively investigated \cite{palmisano2015future}, including spatial frequency \cite{guo2021effects,palmisano1998stimulus}, angular velocity \cite{allison1999effect}, contrast \cite{guo2021effects}, eccentricity \cite{brandt1973differential,andersen1985induced,palmisano1998stimulus,allison1999effect}, and the role of motion parallax \cite{palmisano2004jitter,palmisano2011simulated,palmisano2012horizontal,seno2017oscillating}. As a result, the stimulus features that effectively induce vection are well characterized, as vection tends to occur with visual stimuli that have low spatial and temporal frequency, small angular velocity, high contrast, and the presence of attentional distractors \cite{seno2009}.

While vection typically produces the sensation of self-motion opposite to the direction of the visual motion, the present study focuses on a phenomenon in which the body actually moves in the same direction as the visual motion. This phenomenon is known as postural drift, or visually induced postural response (VIPR) \cite{hanssens2013visually,sugiura2017experimental}.
Vection and VIPRs are often closely related. Large-field visual motion can induce an illusory sense of self-motion (vection), and the onset and magnitude of vection tend to change together with visually induced body sway \cite{thurrell2002vection, kawakita2000body, guerraz2008mechanisms, kunihiro2004postural}. Under these conditions, large-field visual motion often leads observers to tilt or shift their body in the direction of the visual motion \cite{dichgans1978visual, kunihiro2004postural}. 
Using this visuo-postural coupling, our study attempts to control pedestrian trajectories with visual motion stimuli.

\subsection{Pedestrian Control}



In public spaces, pedestrian flow is typically guided by signage---such as text, symbols, or pictograms \cite{zwahlen1999legibility,fuller2002arrow}. Dynamic, animated signage has also been proposed \cite{galea2014experimental,galea2017international,galea2017evaluating}. In addition, physical and compulsory measures (e.g., belt partitions) as well as human-operated guidance (e.g., verbal instructions from police officers or staff) are commonly used. These approaches rely on conscious attention and interpretation, and thus provide strong and explicit means of directing pedestrian flow. %
However, extensive reliance on visually salient signage can make urban spaces visually cluttered and cause important information to be overlooked unless it is viewed within central vision \cite{jamshidi2020wayfinding}, and human-operated guidance requires substantial staffing resources. %
By contrast, the MDS-based approach explored in this study has the potential to modulate pedestrian flow in a more implicit, nonlinguistic manner while reducing the need for human labor.

Several studies have explored nonverbal methods that guide pedestrian behavior using additional infrastructure or wearable devices. A method relevant to our work uses monochrome laterally moving stripes on the floor to modulate pedestrian trajectories \cite{furukawa2011vection,sakamoto2019guided}. The effectiveness of these floor-based stimuli has been demonstrated with projection mapping and VR head-mounted displays. One practical implementation is a power-free system where grayscale stripes are printed on the floor and overlaid with lenticular lenses, allowing pedestrians to perceive motion and, as a result, vection simply by walking over them. Other environment-based, non-wearable methods, such as mid-air haptics \cite{suzuki2019midair} and spatial auditory cues \cite{strachan2005gpstunes}, have also been proposed for pedestrian guidance. Other studies have proposed handheld or wearable guidance systems, including smartphone navigation systems \cite{arikawa2007navitime}, head-mounted displays \cite{ishii2016optical,sousa2009head}, and haptic interfaces \cite{tsukada2004activebelt,gomi2019innovative,gleeson2009communication}. 

In contrast to prior approaches, our research is motivated by the possibility of leveraging MDS already installed in urban environments, without requiring additional hardware installations or wearable devices. Accordingly, this study explores whether such existing MDS can be used to influence natural pedestrian flow in public spaces.

\section{Visual Stimuli Design}\label{sec:vs-design}

In this section, we describe the MDS setup (Sec.~\ref{sec:ds-placement}) and the visual-motion stimuli presented on the MDS (Secs.~\ref{sec:strip-params} and \ref{sec:adv}), which were designed based on stripe parameters previously shown to elicit strong vection and were common to both subsequent experiments (Secs.~\ref{sec:exp1} and \ref{sec:field-exp}).

\subsection{MDS setup}\label{sec:ds-placement} 

The MDS setup was common to both experiments. 
The MDS comprised six digital signage displays (DSs), arranged in two columns (lateral direction) and three rows (depth direction). Each DS was vertically oriented, and the center-to-center spacing between adjacent DSs was 3.5 m in both directions. Although this spacing was slightly narrower than that typically used in public spaces such as railway stations (approximately 4–6 m), it was chosen to accommodate the spatial constraints of both experimental environments.

\begin{figure}[t]
\centering
  \includegraphics[width=0.6\linewidth]{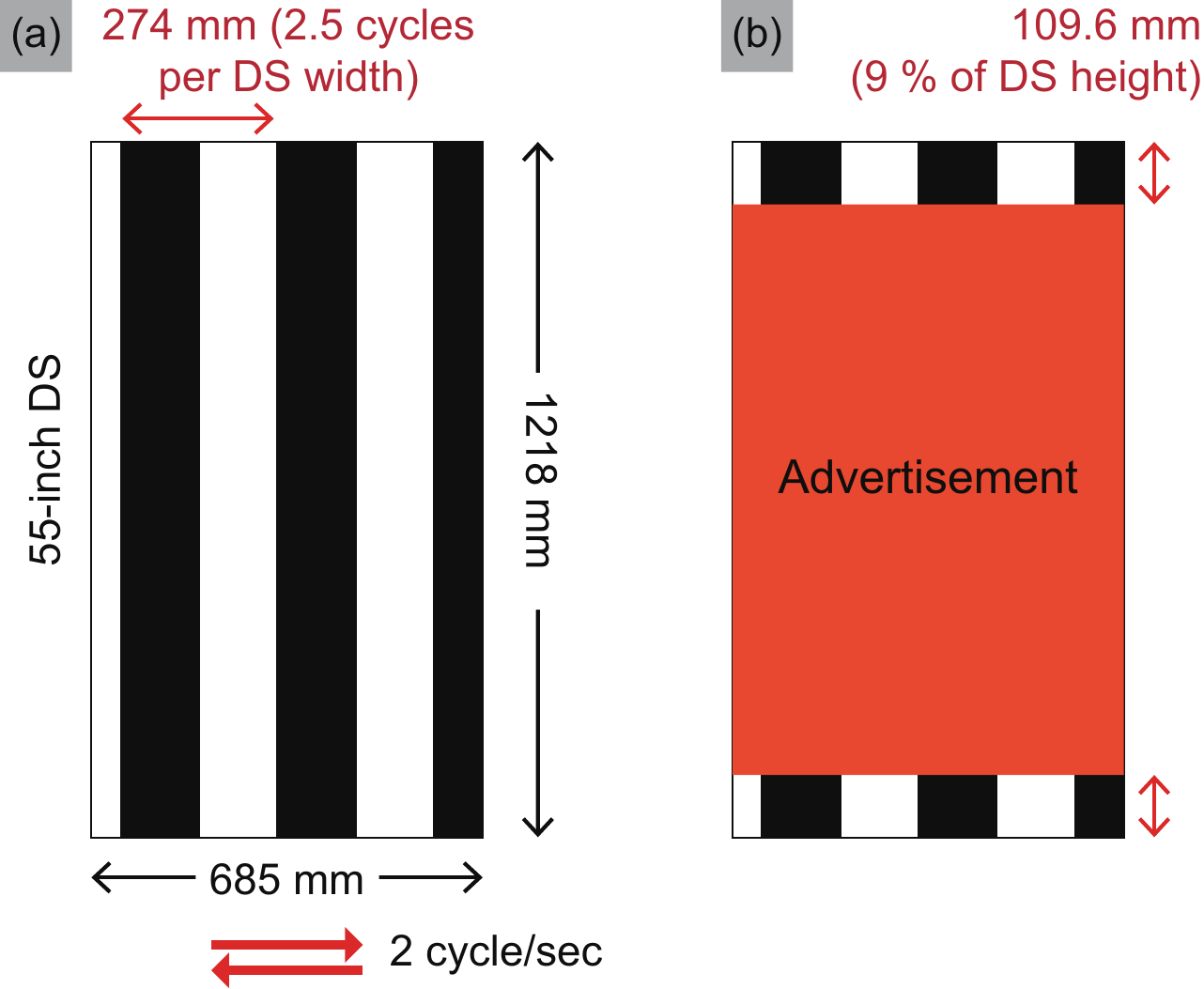}
    \caption{Overview of visual stimuli common to Experiments 1 and 2 (Secs~\ref{sec:exp1} and \ref{sec:field-exp}). 
    (a) Full-screen stripe pattern presented on the MDS, used on Day 1 of Experiment 1 and on both days of Experiment 2. (b) Partial stripe pattern presented on the MDS in regions above and below the advertisements, used only on Day 2 of Experiment 1.}
  \label{fig:common-env}
\end{figure}

\subsection{Stripe Design}\label{sec:strip-params}

The visual-motion stimuli used in this study were based on laterally moving stripe patterns, as illustrated in Figure~\ref{fig:teaser}. Previous vection research has shown that effective stimuli typically exhibit the following characteristics \cite{seno2009}: low spatial frequencies of approximately 0.045–0.286 cycles/deg (= 3.5–22.2 $^\circ$/cycle) \cite{yu2010perception}, low temporal frequencies of 2–7 Hz \cite{sauvan1995spatiotemporal}, slow angular velocities of 5–30 $^\circ$/sec \cite{sauvan1993properties,riecke2011compelling,wang2021visually}, high contrast \cite{yu2010perception,guo2021effects}, presentation in the peripheral visual field \cite{brandt1973differential,palmisano1998stimulus}, and figure–ground organization that supports motion perception \cite{kitazaki2003attentional}.

Following these findings, we designed full-screen, black-and-white stripes with 2.5 cycles across the width of a vertical DS, as illustrated in Figure~\ref{fig:common-env}(a). For a 55-inch DS (display area: 686 mm × 1218 mm), this corresponds to a stripe cycle width of 274 mm. The temporal frequency of the stripe motion was set to 2 Hz.

\begin{table}[t]
  \centering
  \small
  \caption{Stripe parameters for the front, central, and rear DS pairs at a viewing distance of 2.0 m from the front DSs}
  \label{tbl:strip-params}
  \begin{tabular}{ccccc}
  \tiny
    & Viewing 
    & Angular cycle  
    & Angular & \multirow{2}{*}{Eccentricity [$^\circ$]}\\
    & distance {[m]}
    & width {[$^\circ$/cycle]}  
    & velocity {[$^\circ$/sec]} 
    & \\
    \midrule
    Front & 2.0 & 7.87 & 15.9 & 35.1-46.3 \\
    Central & 5.5 & 2.86 & 5.72 & 14.4-20.8 \\
    Rear & 9.0 & 1.74 & 3.49 & 8.9-13.1 \\
  \end{tabular}
\end{table}
Assuming the MDS layout and the initial observer position in the laboratory experiment (2 m away from the front DS pair; see Figure~\ref{fig:env-WS}(a)), the angular cycle width, angular velocity, and eccentricity of the stripe pattern for each of the three DS pairs are summarized in Table~\ref{tbl:strip-params}. The front DSs, in particular, satisfy the recommended parameter ranges for effectively inducing vection. Although the central and rear DSs have lower values across these parameters at the initial viewing position, pedestrians approach each DS as they walk through the space; their viewing distance therefore decreases and the corresponding stimulus parameters increase, making the visual conditions progressively more favorable for strong vection. 
When the observer fixated straight ahead from the initial position (2.0 m from the front DS pair), eccentricity ranged from 8.9$^\circ$ to 46.3$^\circ$ (Table~\ref{tbl:strip-params}), indicating that the MDS images primarily stimulated the peripheral visual field.

Visual stimuli containing multiple depth layers have also been shown to enhance vection \cite{palmisano2004jitter,palmisano2011simulated,palmisano2012horizontal,seno2017oscillating}. In our study, the MDS layout consists of three depth layers—front, central, and rear pairs of DSs. This configuration may enhance vection by exploiting the unique characteristics of MDS as spatially separated displays, which fundamentally differ from a single continuous screen.

\subsection{Advertisement-Compatible Partial-Stripe Design}\label{sec:adv}

Because contemporary MDS installations are primarily used for advertising, a practical implementation should remain compatible with existing advertisements. To this end, we propose presenting laterally moving stripe patterns above and below the advertisement area, as shown in Figure~\ref{fig:common-env}(b). The stripe parameters follow those in Sec.~\ref{sec:strip-params}. We adopt a practical layout consistent with standard advertisement formats: stripe regions occupying 9\% of the DS height (109.6 mm) at both the top and bottom, yielding an advertisement aspect ratio of approximately 3:4.37, close to the standard 3:4 format.

This design not only maintains compatibility with existing advertisements but also enhances figure–ground perception \cite{kitazaki2003attentional}; the advertisement is likely perceived as the figure and the stripes as background, which may facilitate vection compared with a simple full-screen stripe presentation. However, because the angular extent of the motion stimulus is substantially smaller than in the full-screen configuration, a reduction in vection strength is also expected \cite{nakamura2006effects}.

\section{Experiment Design}\label{sec:exp-design}

In this study, two experiments were conducted. Experiment 1 was a controlled laboratory experiment (Sec.~\ref{sec:exp1}) in which participants were instructed to walk individually toward a designated destination. This experiment tested both the full-screen stripe pattern (Sec.~\ref{sec:strip-params}, Figure~\ref{fig:common-env}(a)) and the partial stripe pattern designed to be compatible with advertisements (Sec.~\ref{sec:adv}, Figure~\ref{fig:common-env}(b)), using a within-subjects design for each day. The objective was to examine whether visual motion presented across the MDS setup could modulate pedestrian trajectories in the direction expected based on prior vection research under controlled laboratory conditions.

Experiment 2 was a field experiment (Sec.~\ref{sec:field-exp}) conducted in a national museum, where a large number of visitors walked freely without experimental control. This experiment examined only the full-screen stripe pattern (Sec.~\ref{sec:strip-params}, Figure~\ref{fig:common-env}(a)). It was conducted over two days, with the direction of stripe motion reversed across days. 
Building on the findings of Experiment 1, the objective was to investigate whether comparable direction-dependent differences would also emerge in aggregate pedestrian positions during unconstrained, naturalistic pedestrian flow in a public space.

The visual stimulus design and spatial layout of the MDS were identical in both experiments, as described in Sec.~\ref{sec:vs-design} and illustrated in Figure~\ref{fig:common-env}.

\section{Experiment 1: Laboratory Experiment}\label{sec:exp1}


A laboratory experiment was conducted to evaluate the effects of the proposed visual-motion stimuli on pedestrian trajectories using MDS under controlled conditions (designated starting positions and destinations, and single-pedestrian walking only).

\begin{figure*}[t]
  \includegraphics[width=\linewidth]{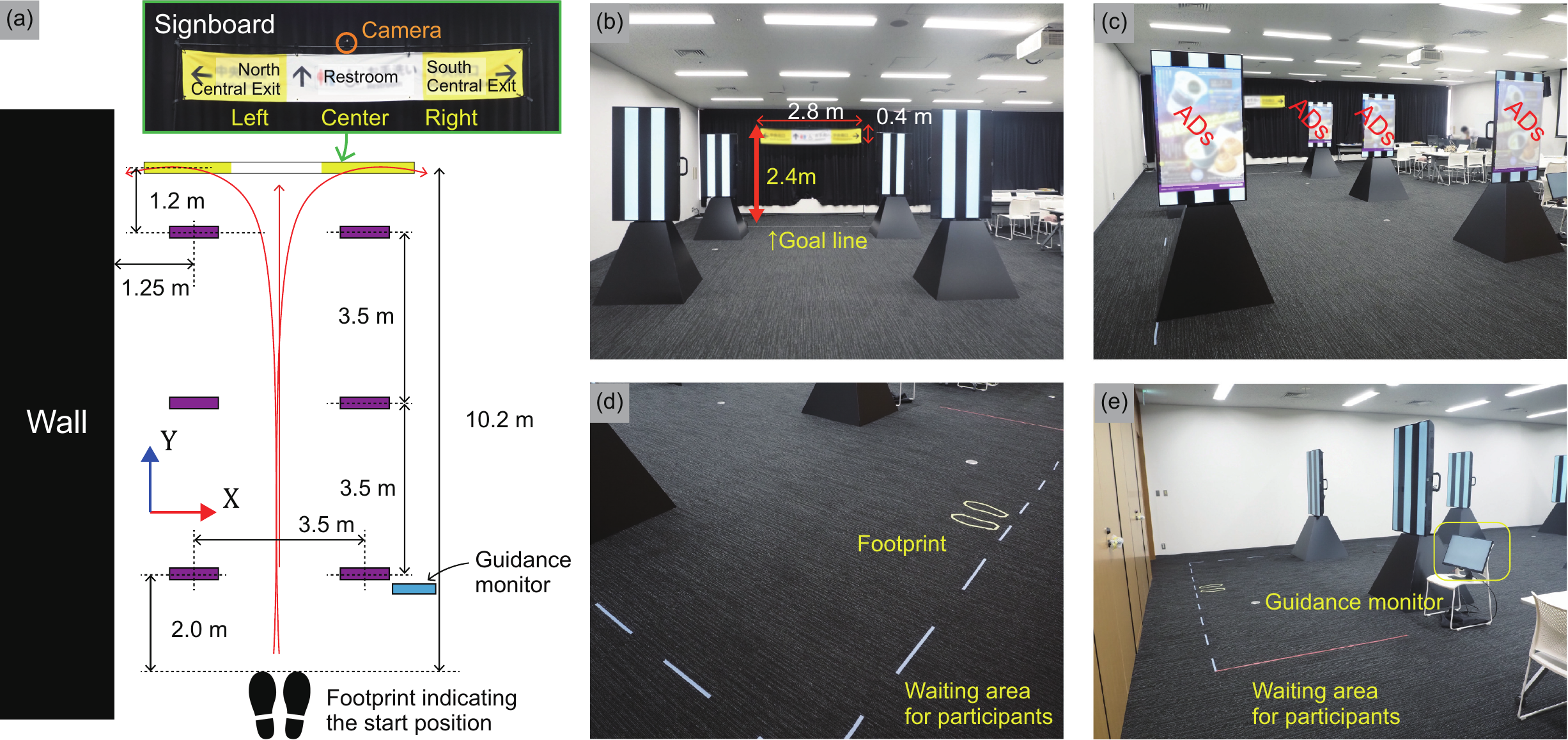}
  \caption{(a) Spatial layout for Experiment 1. A station-like signboard was suspended behind the MDS, indicating three destinations. (b) Photograph of the space from the front when the full-screen stripe pattern was displayed on the MDS (the two front DSs are out of frame). (c) Photograph from the front-left when the partial stripe pattern, compatible with existing advertisements, was displayed on the MDS. (d) Footprints indicating the starting position. (e) Monitor displaying each participant’s designated destination.}
  \label{fig:env-WS}
\end{figure*}

\subsection{Environment}




The experimental setup is shown in Figure~\ref{fig:env-WS}(a). Six 55-inch, vertically oriented DSs (JAPANNEXT JN-IPS5503TUHDR, 1080×1920 px, display area: W684.9 mm × H1217.6 mm) were arranged with a center-to-center spacing of 3.5 m in both the horizontal and depth directions. The center height of the MDS array was 1.4 m. The experiment was conducted in a 160 m$^2$ indoor conference room, and the rear windows were covered with black curtains to block outdoor light (Figure~\ref{fig:env-WS}(b)). 

As shown in Figure~\ref{fig:env-WS}(a), 
the experimental space was not left-right symmetric: the left side was bounded by a wall, whereas the right side was open. This asymmetry is a limitation of Experiment 1 and should be considered when interpreting absolute trajectory values and differences between left and right destinations. Because the geometry was fixed across all trials, however, relative differences among stripe-motion conditions were evaluated under the same spatial constraint.

The starting position was indicated by a footprint marker (Figures~\ref{fig:env-WS}(a) and \ref{fig:env-WS}(d)) placed 2 m in front of the front DSs. All participants were instructed to align their toes with the top of this marker, and staff members ensured correct positioning. The goal line, marked with white tape, was set 1.2 m behind the rear DSs.

To simulate walking toward a station exit or transfer gate in a public space, a signboard mimicking a local railway sign (W2.8 m × H0.4 m) was suspended at 2.4 m above the floor (Figures~\ref{fig:env-WS}(a) and \ref{fig:env-WS}(b)). The left side of the signboard indicated “North Central Exit,” the center “Restroom,” and the right side “South Central Exit,” each accompanied by directional arrows. These labels were written in the native language of the country in which the experiment was conducted. 

Participants’ walking trajectories were recorded using an RGB–Depth camera (Orbbec 335L; 30 fps; 1280×720 px; stereo). The camera was mounted above the signboard, as shown in the upper right of Figure~\ref{fig:env-WS}(a), to capture the entire MDS space.

\subsection{Participants}\label{sec:exp1-participants}

\begin{figure}[t]
\centering
  \includegraphics[width=\linewidth]{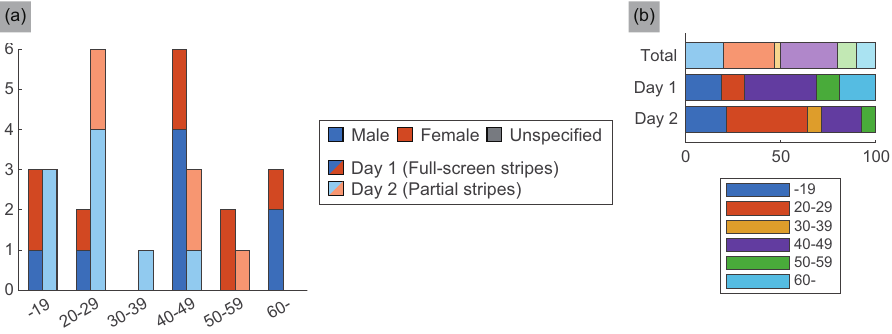}
  \caption{Participant age and gender distribution in Experiment 1 (Sec.~\ref{sec:exp1}). (a) Number of participants by age group, gender, and day. (b) Proportion of participants by age group and day.}
  \label{fig:age-range}
\end{figure}

Participants were attendees of participatory workshops held in a national museum. Day 1 had 16 participants (8 female), Day 2 had 14 (5 female), and no one attended on both days.

Participants reported their ages in decade-based categories. The age and gender distributions and proportions are shown in Figure~\ref{fig:age-range}. As participants were recruited without age or gender restrictions, the demographic distribution varied between days: participants in their 40s dominated Day 1, whereas those in their 20s accounted for about 40\% on Day 2. On both days, all participants in their 50s were female, and only one participant was in their 30s. No participant chose the “other” option for gender. Nevertheless, we consider the study successfully recruited participants across a broad range of age groups and genders.

Participants were required to have adequate binocular visual acuity (with glasses or contact lenses if necessary) and normal walking ability. All participants were naïve to the purpose of the experiment. 
The experimental protocol was approved by the research ethics committee of NTT, Inc. and conducted in accordance with the ethical standards of the Declaration of Helsinki (this also applies to the subsequent experiments). Written informed consent was obtained from all participants prior to the experiment.




\subsection{Stimuli}\label{sec:exp1-stimuli}
On Day 1, participants viewed a full-screen stripe pattern (Figures~\ref{fig:common-env}(a) and \ref{fig:env-WS}(b)). In contrast, on Day 2, they viewed a partial stripe pattern designed to be compatible with an existing advertisement (Figures~\ref{fig:common-env}(b) and \ref{fig:env-WS}(c)). 
The luminance of the DS was 352.0 cd/m$^2$ for white and 0.31 cd/m$^2$ for black, as measured with a colorimeter (KONICA MINOLTA Ltd., CS-150).

\subsection{Condition}

On Day 1, stripe motion had three levels (rightward, leftward, stationary), and destination also had three levels (right: ``South Central Exit,'' center: ``Restroom,'' left: ``North Central Exit''; see Figure \ref{fig:env-WS}(a)), resulting in 3×3 = 9 conditions. These full-screen stripe conditions were used to characterize the basic properties of the effect. On Day 2, stripe motion again had three levels (rightward, leftward, stationary), while destination had two levels (right: ``South Central Exit,'' left: ``North Central Exit''), resulting in 3×2 = 6 conditions. To reduce participant burden, the partial-stripe stimuli were tested only in the conditions necessary to evaluate the effect of stripe-motion direction. Each participant completed one trial per condition, resulting in nine passes through the signage space on Day 1 and six on Day 2.

%

To facilitate smooth experimental operation, the stripe-motion conditions were presented in blocks shared by all participants. The visual stimulus on the MDS remained unchanged while all participants completed the trials scheduled under the current stripe-motion condition and was updated only between blocks. Within each stripe-motion block, the destination was displayed separately for each participant on the guidance monitor immediately before each trial, and the order of destinations was randomized independently for each participant. Each participant experienced every combination of stripe-motion and destination conditions exactly once.



\subsection{Procedure}

\begin{figure}[t]
  \includegraphics[width=\linewidth]{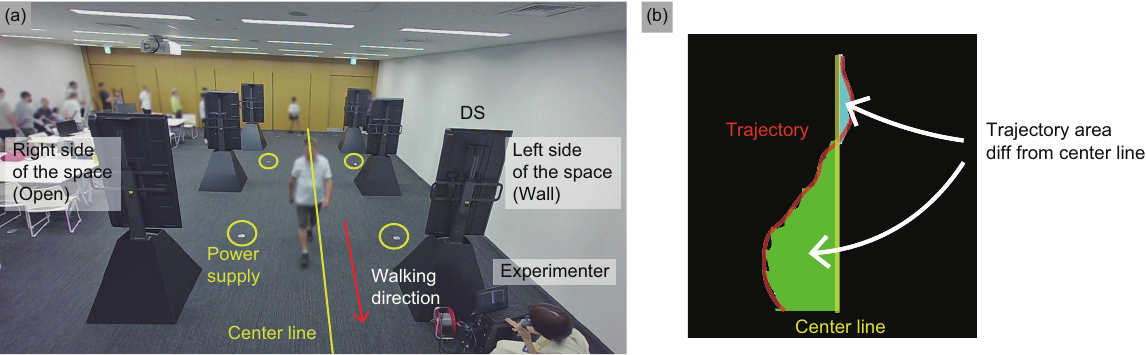}
    \caption{(a) RGB camera image captured from a camera mounted above the signboard during the experiment. Participants are blurred to protect privacy. (b) Using four known reference points shown in (a) (power supplies) and the RGB camera parameters, participants’ trajectories were projected onto the floor $X-Y$ coordinate system, and the area representing deviation from the central line was computed as a summary statistic for each trajectory.}
  \label{fig:analysis-WS}
\end{figure}

The workshop lasted 1.5 hours per session, and the walking experiment was conducted during the first 30 minutes.

Prior to the experiment, each participant was randomly assigned a number determining their walking order, which remained fixed throughout the experiment. The walking procedure for each trial was as follows. Before starting to walk, participants aligned their toes with the footprint markers shown in Figures~\ref{fig:env-WS}(a) and~\ref{fig:env-WS}(d). As shown in Figure~\ref{fig:env-WS}(e), a monitor positioned near the starting location displayed one of three destinations depicted on the signboard, randomly specifying the destination for each trial (a staff member standing near the monitor also verbally indicated the destination). Participants then walked toward the designated destination until they crossed the goal line, after which they returned to the entrance from either the left or right side of the space. Those with central or right destinations returned from the right side, and those with left destinations returned from the left side.

Each participant completed a single practice walking trial before the main experiment.




\subsection{Analysis}\label{sec:exp1-analysis}


RGB and depth images were processed in Python 3.11.8 using YOLO (v8.3.178) for pedestrian position detection. As illustrated in Figure~\ref{fig:analysis-WS}(a), a bounding box was detected around each person in every frame, and the midpoint of its bottom edge was recorded as the image coordinate of the pedestrian’s footpoint, $\bm{P}_i$. Using the known positions of four reference points in the image (white circles indicating power supplies) and the RGB camera parameters, $\bm{P}_i$ was transformed into spatial coordinates $\bm{P}_s = (X, Y, Z)$. Assuming that the footpoint lies on the floor ($Z=0$), the resulting $X$–$Y$ coordinates were treated as trajectories on the floor plane (Figure~\ref{fig:analysis-WS}(b)).
Because the goal line lay outside the camera’s field of view (Figure~\ref{fig:analysis-WS}(a)), the analysis was restricted to a 10.0 m depth range from the starting position, although the actual distance from start to goal was 10.2 m (see Figure \ref{fig:env-WS}(a)).

The summary statistic for each trajectory was defined as the area of deviation from the central line (Figure~\ref{fig:analysis-WS}(b)). To compute this area, the trajectory’s $X-Y$ coordinates of $\bm{P}_s$ were plotted onto an image using OpenCV 3.4.12, with the coordinates scaled by a factor $s$ so that they fit within the image (red line in Figure~\ref{fig:analysis-WS}(b)), together with the central line of the MDS space (yellow line in Figure~\ref{fig:analysis-WS}(b)). Contours enclosed by the trajectory and the central line were then detected, and the area of each contour was calculated. Areas on the left side of the central line (from the pedestrian’s perspective, the right-hand side) were assigned positive values, and those on the right were assigned negative values. The sum of these signed areas was used as the summary statistic for each walking trajectory and was then rescaled to the original coordinate system by multiplying by $1/s^2$.


\subsection{Results and Discussion}





\begin{figure}[t]
\centering
  \includegraphics[width=0.7\linewidth]{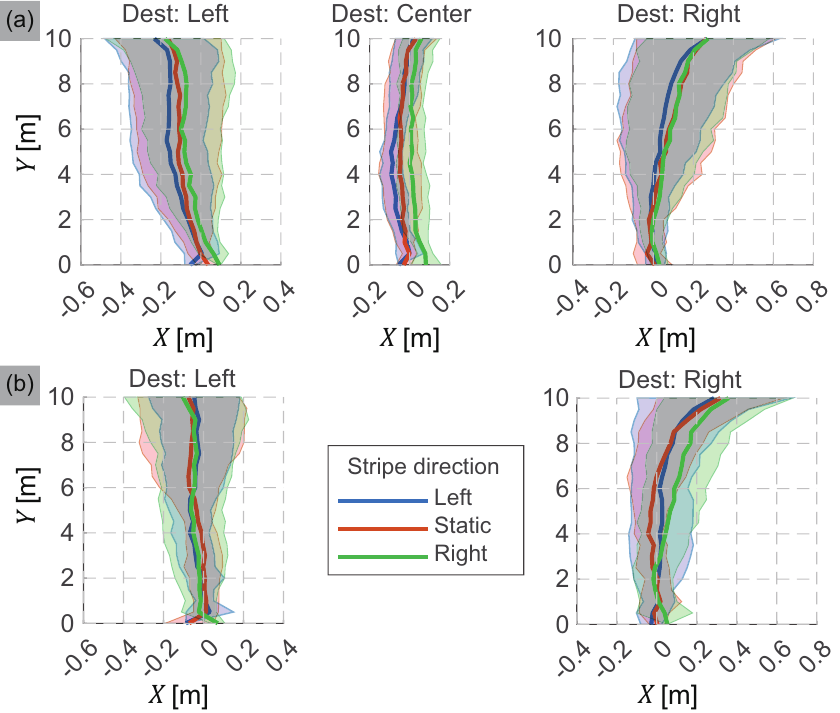}
    \caption{Graphs of participants' $X$-axis displacement, computed every 0.05\,m along the walking direction ($Y$-axis), for each day and condition. (a) Day~1 (full-screen stripe pattern). (b) Day~2 (partial stripe pattern). The shaded regions indicate $\pm1$\,SD in $X$.}
  \label{fig:res-WS}
\end{figure}

Figure~\ref{fig:res-WS} shows the mean and standard deviation (SD) of walking trajectories along the $Y$-axis for each experimental condition: (a) the Day 1 full-screen stripe pattern and (b) the Day 2 partial stripe pattern. The curves were obtained by sampling at 0.05 m intervals along the walking direction ($Y$-axis) and, at each sample, computing the mean and SD of all points within $\pm 0.25$ m in $Y$.
Note that regions farther from the RGB camera were captured at lower resolution (up to approximately 2 cm per pixel near the starting position), and the bottom-edge midpoint of the YOLO bounding box did not necessarily coincide with the participants' actual foot positions, leading to larger detection errors around the start position. As a result, in some graphs, the starting position ($y = 0$) does not precisely align with $x = 0$.

\begin{figure}[t]
\centering
  \includegraphics[width=0.7\linewidth]{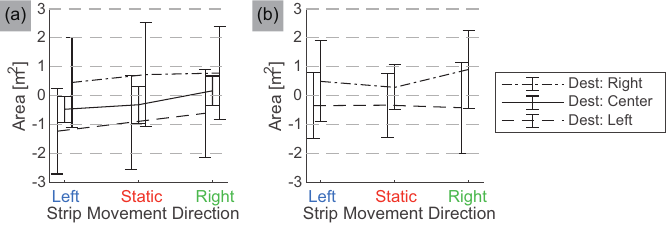}
    \caption{Mean values of trajectory areas for each condition on (a) the full-screen stripe pattern and (b) the partial stripe pattern, respectively. Error bars represent $\pm$1 SD.}
  \label{fig:res-WS-area-stat}
\end{figure}

Figure~\ref{fig:res-WS-area-stat} shows the means and SDs of the trajectory area, which is defined as a difference area from the central line (see Figure~\ref{fig:analysis-WS}(b)), for each condition. %
All the raw trajectory data of each participant and each condition are shown in the Supplementary Material (Figures S1–S4).

\subsubsection{Analysis on Day 1 (full-screen stripe pattern)}\label{sec:analysis-Day1}
A two-way repeated-measures ANOVA was performed on the trajectory area. 
Statistical analyses were performed using MATLAB R2025b; the same software was used for all subsequent statistical analyses.
Mauchly’s tests indicated that the sphericity assumption was satisfied for stripe motion ($W = 0.829$, $\chi^2(2) = 2.63$, $p = 0.269$) and for the interaction between stripe motion and destination ($W = 0.616$, $\chi^2(9) = 6.51$, $p = 0.688$), but was violated for destination ($W = 0.115$, $\chi^2(2) = 30.31$, $p < 0.001$). A Greenhouse–Geisser correction was therefore applied only to the main effect of destination ($\epsilon = 0.530$). 

The analysis revealed significant main effects of stripe motion, $F(2, 30) = 14.49$, $p < 0.001$, $\eta_p^2 = 0.491$, and destination, $F(1.06, 15.91) = 4.63$, $p = 0.045$, $\eta_p^2 = 0.236$. The interaction between stripe motion and destination was not significant, $F(4, 60) = 0.88$, $p = 0.481$, $\eta_p^2 = 0.055$.

Post-hoc pairwise comparisons using the Tukey–Kramer procedure indicated significant differences between the left and right stripe-motion conditions ($p < 0.001$) and between the static and right conditions ($p = 0.046$), whereas the difference between left and static was not significant ($p = 0.075$). The same post-hoc analyses revealed no significant pairwise differences among the destination conditions. The absence of a significant difference between the left and static stripe-motion conditions may be attributable to a wall on the left side of the space, which likely constrained participants' ability to deviate or walk dynamically toward that side.


\subsubsection{Analysis on Day 2 (partial stripe pattern)}


A two-way repeated-measures ANOVA was performed on the trajectory area. 
Because Mauchly’s tests indicated violations of the sphericity assumption for stripe motion ($W = 0.430$, $\chi^2(2) = 10.13$, $p = 0.006$) and the interaction between stripe motion and destination ($W = 0.383$, $\chi^2(2) = 11.53$, $p = 0.003$), Greenhouse–Geisser corrections were applied to these effects ($\epsilon = 0.637$ and $0.618$, respectively). No sphericity correction was required for destination because it had only two levels.

The analysis revealed a significant main effect of destination, $F(1, 13) = 7.13$, $p = 0.019$, $\eta_p^2 = 0.354$. In contrast, the main effect of stripe motion was not significant, $F(1.27, 16.56) = 0.32$, $p = 0.632$, $\eta_p^2 = 0.024$. The interaction between stripe motion and destination was also not significant, $F(1.24, 16.08) = 1.60$, $p = 0.229$, $\eta_p^2 = 0.109$.
These statistical results are consistent with the plots shown in Figures~\ref{fig:res-WS}(b) and \ref{fig:res-WS-area-stat}(b). 


The absence of a significant directional effect suggests that the limited angular extent of the partial-stripe stimulus, which was designed to accommodate advertisements, may have reduced its capacity to modulate pedestrian trajectories, while destination remained a significant determinant of walking behavior.
This result is also consistent with prior fundamental studies of vection, which have argued that larger presentation areas produce stronger vection effects \cite{nakamura1998stimulus,tamada2015roles}.

\subsection{Summary of Experiment 1}

In a controlled experiment conducted during a participatory workshop at a national museum, full-screen stripe patterns displayed on the MDS significantly modulated pedestrian trajectories across a diverse range of ages and genders, whereas no significant directional effect was detected for partial-stripe motion in a separate group of participants. In both conditions, the destination effectively modulated walking trajectories. These findings demonstrate the feasibility of behavioral modulation via MDS and suggest that practical implementations may require a sufficiently large motion area. Rather than confining narrow moving stripes to a small portion of the display to avoid disrupting the existing advertisement, more extensive motion effects—such as animating the advertisement itself or embedding subtle motion into its background—may be needed to produce measurable behavioral modulation in practical applications.

\section{Experiment 2: Field Experiment in a National Museum}\label{sec:field-exp}

The field experiment was conducted over two days in a free-admission area of a national museum to examine whether the aggregate pedestrian positions differed between two field-test days presenting opposing directions of visual motion on the MDS under highly unconstrained conditions (free starting positions, unspecified destinations, and walking in groups).

The experiment was conducted during a two-day event expected to attract a large number of visitors. To ensure a balanced comparison between conditions, the evaluation was limited to leftward and rightward full-screen stripe motion.

\subsection{Environment}



\begin{figure}[t]
\centering
  \includegraphics[width=\linewidth]{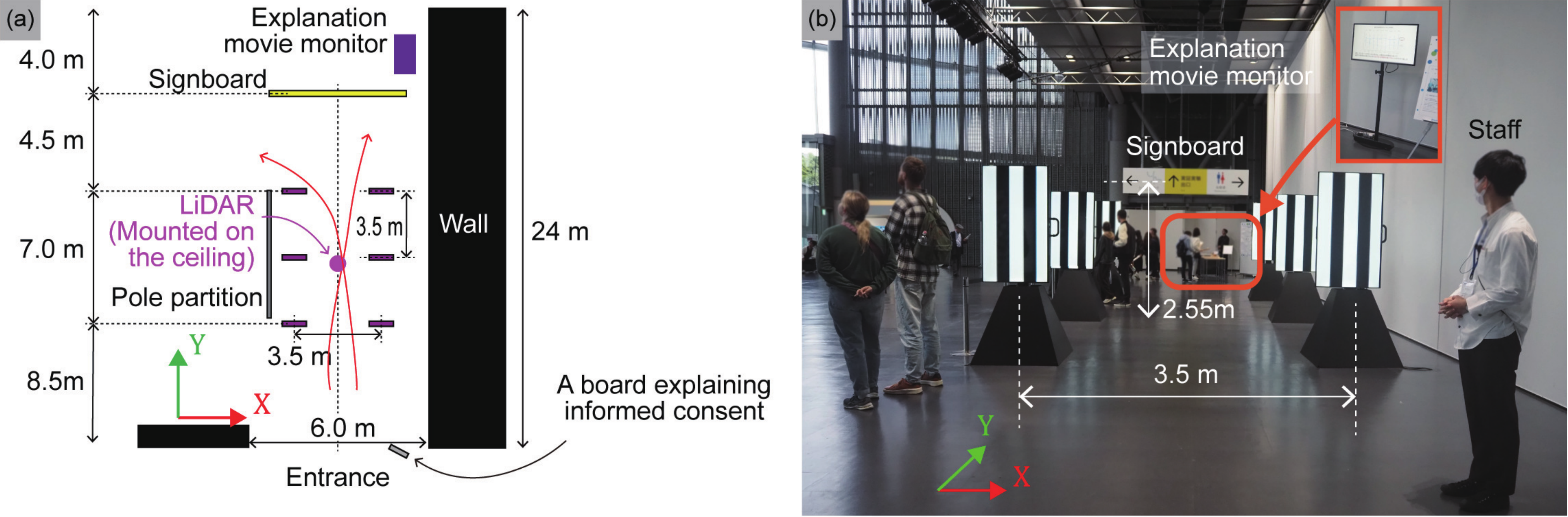}
    \caption{(a) Layout of the field-experiment space (Experiment 2) in a free-admission area of a national museum. Unlike Experiment 1, the right side was bounded by a wall while the left side remained open; the approach path was longer, and an explanation booth with a movie monitor was installed beyond the MDS. (b) Front view of the experiment space. Agency staff stood at the front-right corner to invite pedestrians, and a station-like signboard was suspended behind the MDS units.}
  \label{fig:env-FT}
\end{figure}

Six 55-inch, vertically oriented DSs, identical to those used in Experiment 1, were installed in the layout shown in Figure~\ref{fig:env-FT}(a). The MDS was open on the left side and bounded on the right by a wall located 1.75 m lateral to the rightmost DS. Such a layout was inevitable given the original spatial design of the free-admission area and the circulation plan intended to accommodate a large number of visitors passing through from the entrance. The distance from the entrance to the first DS was 8.5 m, and the distance from the rear DSs to the signboard was 4.5 m. A custom-made signboard (3.5 m wide, 0.7 m high) was suspended with its center 2.55 m above the floor. As in Experiment 1, the signboard, resembling a station sign, indicated three possible destinations within the space: other exhibitions to the left, the explanation movie monitor straight ahead, and the restrooms to the right.

To measure pedestrian flow, a LiDAR sensor (Hesai Technology JT128) with a hemispherical field of view was suspended from the ceiling at approximately 4.5 m above the floor. 
An explanation booth with a movie monitor for the experiment was installed along the wall in the rear space of the MDS (see Figure~\ref{fig:env-FT}(a)), as it was required to provide visitors with a detailed explanation of the research after they passed through the MDS space in a museum. As shown in Figure~\ref{fig:env-FT}(b), the booth was inevitably visible from the entrance, introducing asymmetry in both the spatial layout and the visual field. However, because the arrangement of the space and the booth location were kept constant across both days, comparisons between stripe-motion conditions were made under identical spatial constraints.
The experiment was conducted in a space with large windows, and sound from events in the adjacent area was sometimes present; however, these conditions were similar on both days.


\subsection{Participants}\label{sec:exp2-participants}


All pedestrians who passed through the MDS space during the experimental period were included as participants. Individual written consent forms were not collected because the experiment was conducted as a large-scale public field trial in a free-admission area of a national museum. Instead, a notice-based opt-out consent procedure approved by the institutional ethics committee was used. An information notice was displayed before the MDS space (bottom-right of Figure \ref{fig:env-FT}(a)), informing visitors about the experiment, the data to be collected, and the handling of personal information. Visitors who did not wish to participate could use the adjacent avoidance route. By passing beyond the information notice and entering the MDS space, visitors were regarded as having agreed to participate. The notice also stated that visitors could inform the on-site staff in order to opt out. In such cases, staff members would guide them to an adjacent avoidance route outside the measurement area or arrange for deletion or exclusion of data recorded during the relevant time window. No participant requested exclusion after passing through the space.


According to a staff-operated manual people counter, 1,369 participants passed through the space on Day 1 (rightward stripe motion) and 1,555 on Day 2 (leftward stripe motion). Unlike the controlled laboratory experiment described in Sec. \ref{sec:exp1}, no restrictions were imposed on walking ability or vision. Because pedestrian trajectories were recorded solely using LiDAR, participants’ ages and genders were not individually identified. Nevertheless, visual observation during the experiment suggested that the pedestrian flow included visitors of various age groups. 
Because entry to the free-admission area was not restricted, some visitors may have participated on both days. 
Because participants could not be individually identified, the extent of duplicated participation and its potential influence on the results could not be quantified.


\subsection{Stimuli and Condition}\label{sec:exp2-stimuli}

The visual stimulus was identical to the full-screen stripes used in Experiment 1, consisting of the black-and-white binary stripe pattern shown in Figure~\ref{fig:env-WS}(b). The spatial and temporal frequencies of the stripes were the same as those listed in Sec.~\ref{sec:strip-params}. The stripes moved rightward on Day 1 and leftward on Day 2, and their motion continued throughout the museum’s seven-hour opening period without interruption. 

Alternating the motion direction across shorter time blocks within each day was also considered. However, this approach was deemed inappropriate because visitor arrivals could fluctuate abruptly around the switching times, and visitors who directly observed a reversal in the motion direction might respond differently from those exposed to a continuously presented stimulus. Moreover, because the study aimed to investigate effects on natural pedestrian flows and was conducted as part of a museum experience open to the general public, temporarily preventing visitors from entering around the switching times solely to exclude those who might observe the transitions was considered impractical. The experiment was therefore conducted using a single motion direction on each day. The two experimental days were consecutive public holidays during which the same museum event was held; thus, differences in visitor populations and general museum conditions between the two days were expected to be relatively limited.

This field experiment was conducted as part of a museum exhibition, and the visual stimuli had to be designed to ensure visitors enjoyed the experience. Consequently, it was not possible to include a baseline condition with static images containing no lateral visual motion on the MDS.

\subsection{Procedure}

An agency staff member stood near the entrance to inform visitors about the ongoing field experiment and to invite them to walk through the space. Unlike in the controlled laboratory setting (Experiment 1; Sec.~\ref{sec:exp1}), no restrictions or explicit instructions were imposed regarding starting position or destination; participants simply walked toward their own intended destinations. If a participant stopped midway, staff standing on the left side of the MDS space encouraged them to continue walking. Because the space functioned as an exhibition area, visitors were not prevented from walking in the opposite direction; however, such individuals were excluded from the pedestrian point-cloud data (see Sec.~\ref{sec:exp2-analysis}). No agency staff members entered the MDS space during the experiment.

The experiment was conducted from 10:00 to 17:00, in accordance with the museum’s operating hours, with continuous image presentation on the MDS and continuous point-cloud acquisition by the LiDAR. 

\subsection{Analysis}\label{sec:exp2-analysis}

The point-cloud data in each frame were cropped to the interior region of the MDS space, including an additional 2 m in front of and behind the MDS, resulting in a volume of 2.8 m laterally, 11 m in depth, and 4.5 m in height. This cropping removed non-pedestrian points such as the floor and the MDS, yielding point clouds that primarily represented pedestrians. Pedestrian clustering was performed using DBSCAN \cite{khan2014dbscan}, which provided the centroid of each pedestrian cluster. However, perfect detection was not possible: multiple centroids were occasionally generated for a single person, and the large data volume made manual correction infeasible. Temporal correspondences between centroids in adjacent frames were estimated using the Hungarian algorithm \cite{mills2007dynamic-hugarian}, but occasional failures caused ID swaps, making reliable long-range person-level trajectory reconstruction difficult. Nevertheless, these provisional time-series sequences were used to exclude reverse-walking cases, defined as those whose final point was equal to or behind the starting point.
Through this procedure, centroid data for pedestrians in the space—still containing some noise, including duplicated centroids—were obtained for each day: 294,013 and 313,645 points for Days 1 and 2, respectively.

The original coordinate system for the centroids was defined with its origin at the LiDAR position and the $XY$-plane parallel to the floor. For subsequent analysis, the point-cloud coordinates were transformed as follows: (1) the point cloud was rotated around the LiDAR’s $Z$-axis so that the $Y$-axis aligned with the walking direction and the $X$-axis represented lateral displacement; and (2) a constant offset was applied to the $Y$-coordinates so that the reference line (2 m in front of the front DSs) corresponded to $Y = 0$.

A $t$-test was conducted to examine differences in pedestrians’ spatial distributions between the two days with different stripe directions. Because the temporal correspondence of point clouds was imperfect, individual pedestrians could not be reliably identified. To approximate independence of observations, we (1) computed the median position of all centroids in each frame, yielding a single representative point per frame, and (2) averaged these frame-wise medians over non-overlapping 100-frame intervals (10 s). This procedure collapsed within-frame dependencies, including duplicated centroids for a single person, and reduced temporal autocorrelation, thereby providing approximately independent data points for statistical testing. Given a typical human walking speed of approximately 1.3 m/s \cite{murtagh2021outdoor} and the 11-m longitudinal extent of the experimental space, a 10-s sampling interval corresponded to a walking distance of approximately 12–14 m. Thus, most pedestrians present in the space were expected to have been replaced between successive samples, helping to minimize overlap between observations and preserve their approximate independence.


\subsection{Results}

\begin{figure}[t]
\centering
  \includegraphics[width=0.8\linewidth]{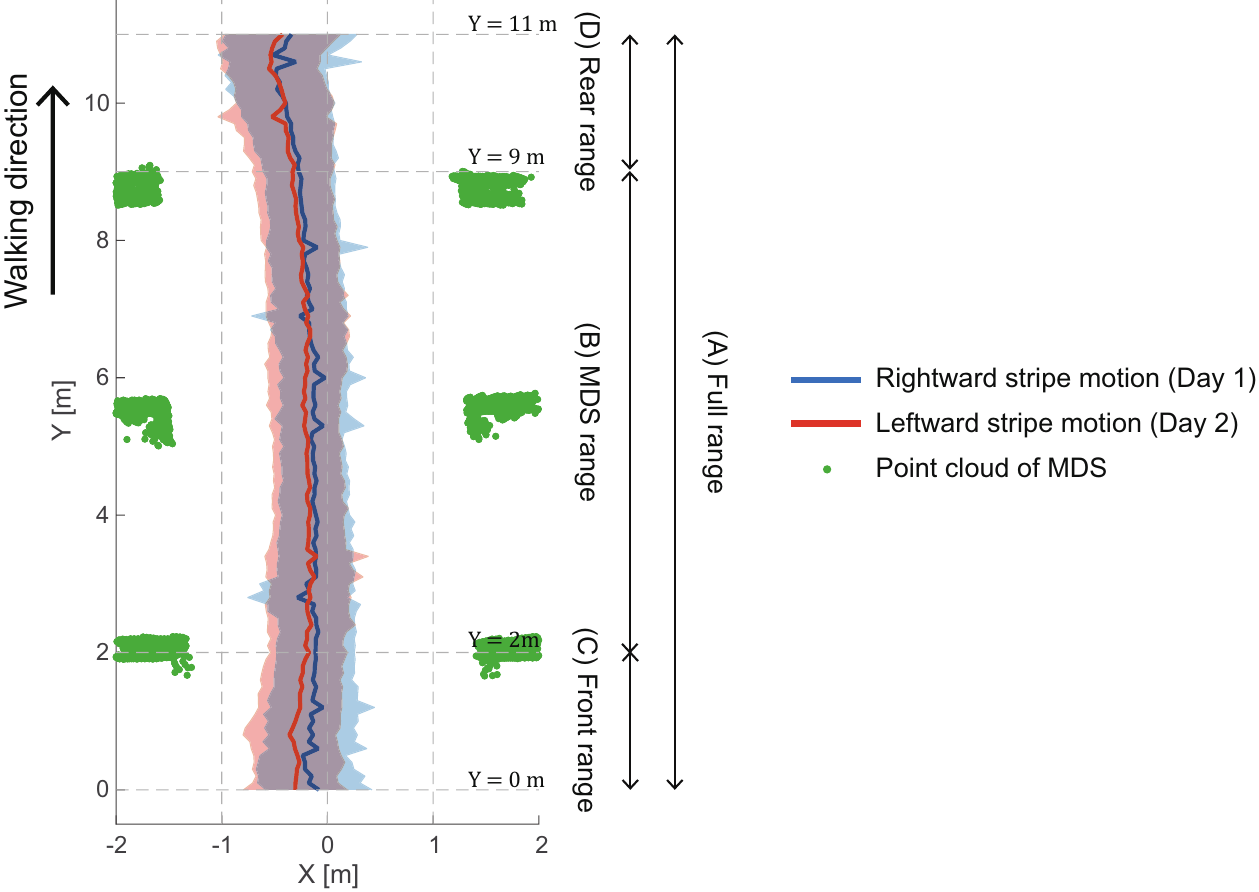}
    \caption{Results of the field test (Experiment 2) showing pedestrians’ spatial distributions on Days 1 and 2. Along the walking direction ($Y$-axis), data were sampled at 0.1 m intervals; for each sample, all points within $\pm 0.05$ m in $Y$ were used to compute the mean and standard deviation (SD) of the $X$ coordinates. Note that the data used to compute these means and SDs are frame-wise median positions across multiple pedestrians.}
  \label{fig:res-FT}
\end{figure}

\begin{table}[t]
  \centering
  \caption{Results of the statistical analysis, including the $t$-test, for the field experiment in the national museum (Experiment 2) for four types of space sections. $\Delta X$ was computed by averaging $X$ coordinates across all the representative (median) coordinates of each frame, indicating the difference in the mean lateral displacement between Days 1 and 2. The $t$-test statistic in the right columns corresponds to the data groups obtained by averaging these frame-wise representative values (medians) every 100 frames.}
  \label{tbl:res-FT}
  \small
 \begin{tabular}{llcllll}
    \multicolumn{1}{c}{Range} & \textit{Y} range [m] & $\Delta X$ [cm] & \textit{p} & \textit{df} & \textit{t} & Cohen's \textit{d}\\
    \cmidrule(lr){1-1}\cmidrule(lr){2-2}\cmidrule(lr){3-3}\cmidrule(lr){4-7}
(A) Full  & $0 \leq Y \leq 11$ & 6.41 & $<$.001 & 1747 & 5.10 & 0.24 \\
(B) MDS  & $2 \leq Y \leq 9$ & 5.44 & $<$.001 & 1095 & 3.91 & 0.24\\
(C) Front  & $0 \leq Y \leq 2$ & 13.11 & $<$.001 & 296 & 5.10 & 0.59\\
(D) Rear  & $9 \leq Y \leq 11$ & 5.91 & 0.017 & 354 & 2.12 & 0.22\\
  \end{tabular}
\end{table}

Figure~\ref{fig:res-FT} shows the mean and standard deviation (SD) of pedestrian positions along the $X$-axis for each condition (each day), obtained by sampling at 0.1 m intervals along the walking direction ($Y$-axis) and, at each sample, computing the mean and SD of all points within $\pm 0.05$ m in $Y$. Note that each data point corresponds to the median position within a frame, as described in the previous section.

Table~\ref{tbl:res-FT} summarizes the statistical analysis for four spatial regions, defined in terms of their \textit{Y}-ranges. The MDS space refers to the area between the front and rear DSs. The Full space comprises the MDS space plus an additional 2-m range along the $Y$-axis both before and after it. The Front and Rear spaces denote the 2-m-long regions along the $Y$-axis immediately before and after the MDS space, respectively.
The mean displacement $\Delta X$ along the $X$-axis for each of the four spatial regions is also summarized in Table~\ref{tbl:res-FT}. In the MDS space, $\Delta X$ was 5.44 cm, compared with 13.11 cm and 5.91 cm in the Front and Rear spaces, respectively. The Front space showed the largest displacement (13.11 cm), whereas the MDS space showed a relatively small one (5.44 cm).

To examine whether the spatial distribution of pedestrians differed significantly between conditions, we performed a one-sided two-sample $t$-test testing the hypothesis that the distribution on Day 1 (rightward stripe motion) was shifted further to the right than the distribution on Day 2 (leftward stripe motion).
The results for each of the four types of spaces are shown in Table \ref{tbl:res-FT}, where the Full, MDS, and Front spaces showed significant differences (all $p < 0.001$), while the Rear space also showed a significant difference ($p=0.017$). 
From the calculated Cohen's $d$, the effect size in the Front space was moderate at $d = 0.59$, while other spaces had relatively small effect sizes at around $d = 0.2$. 
Therefore, the rightward and leftward stripe-motion conditions yielded significant differences in aggregate pedestrian positions in the direction predicted by the visual-motion hypothesis.

\subsection{Discussion}\label{sec:exp2-discuss}


The above analysis suggests that the Front space yielded the most extensive modulation of walking trajectories ($\Delta X=13.11$ cm), which can be reasonably attributed to the fact that all MDS were visible in the Front space. However, the effect within the MDS space itself appears to have been substantially attenuated ($\Delta X = 5.44$ cm) compared to those in the Front space. 
Drawing on informal post-passage interviews, several participants reported deliberately trying not to be “swept away” by the motion of the stripes, or consciously attempting to walk along the central line of the floor (the tiled floor provided a clearly visible central line between the MDS).

These findings suggest that the strong visual-motion images may have elicited a compensatory tendency to resist the induced motion at the cognitive level, thereby reducing lateral displacement within the MDS space. 
These findings indicate that, while the proposed MDS imagery can modulate pedestrian trajectories, there remains a need to develop visual designs with greater public acceptability while maintaining or enhancing their pedestrian-guidance effects.

A relatively small statistical difference in the modulation of pedestrian trajectories in the Rear space may be attributed to the absence of MDS in pedestrians' field of view. However, given that the effect size remained almost unchanged (Cohen’s $d \approx$ 0.2) within and after the MDS space, it is implied that the displacement caused by the vection-inducing images while passing through the MDS space persisted even after passing the MDS space.

Even the largest displacement, $\Delta X = 13.11$ cm in the Front space, seems insufficient for practical pedestrian control or guidance. However, it should be noted that the experimental condition was under predominantly unidirectional flow conditions. In bidirectional pedestrian flow, even a slight laterally induced bias may affect local passing behavior during pedestrian encounters, and these small interaction-level biases could accumulate to stabilize a consistent separation of opposing streams \cite{helbing2001self,zanlungo2012microscopic}. Under such conditions, double-sided MDS may contribute to left- or right-sided pedestrian control. 

\section{Limitations}

Several limitations should be noted. First, the full-screen and partial-stripe stimuli in Experiment 1 were tested on different days with different participant groups, and the stripe-motion conditions were presented in blocks. Therefore, the two types of stripe patterns cannot be directly compared, and possible order effects cannot be fully excluded. 

Second, Experiment 2 assigned one stripe-motion direction to each experimental day. The two conditions were tested on consecutive public holidays during the same museum event, with the spatial and operational settings held constant to minimize day-to-day variation; nevertheless, some residual differences between the days cannot be completely excluded. In addition, a static baseline condition, such as a static full-screen stripe pattern, could not be included because, at the museum’s request, all displayed content had to remain engaging for visitors throughout the exhibition. 
Finally, the LiDAR point-cloud data did not allow reliable tracking of individual pedestrian centroids over extended periods, and manually assigning consistent person IDs across the vast volume of point-cloud data was infeasible. Consequently, the field analysis had to be conducted at the aggregate level, focusing on lateral positional differences rather than individual pedestrian trajectories.

\section{Conclusion and Future Work}

This study investigated whether visual-motion stimuli based on vection research presented on multi-display signage (MDS) in public spaces can modulate pedestrian trajectories. The stimuli consisted of low-spatial-frequency stripe patterns moving laterally, which, in prior basic vection research, have been shown to induce strong vection. We examined two types of stimuli: (1) full-screen stripes (Figure~\ref{fig:common-env}(a)) and (2) partial stripes designed to be compatible with existing advertisements (Figure~\ref{fig:common-env}(b)).

The controlled experiment with participants of diverse ages and genders (Sec.~\ref{sec:exp1}) showed that the full-screen stripes significantly modulated walking trajectories, whereas no significant directional effect was detected for the partial stripes. 
A field experiment (Sec.~\ref{sec:field-exp}) conducted in a national science museum further identified small but direction-consistent differences in aggregate pedestrian positions under opposing full-screen motion conditions. These differences emerged under highly unconstrained conditions, including unrestricted starting positions, unspecified destinations, and group walking, providing strong field-based evidence for the ecological relevance of the proposed approach.

Across both experiments, our findings suggest that effective pedestrian guidance requires reconciling two potentially competing design requirements: visual motion must cover a sufficiently large area to influence walking behavior, while the resulting motion stimuli must remain subtle enough to be socially acceptable—for example, by integrating motion into the advertisement itself or across a large portion of its background.
An important direction for future work is to evaluate the proposed method under bidirectional pedestrian flows, which are common in public spaces. Whereas the present experiments were limited to predominantly unidirectional walking, future studies should examine whether the method can promote consistent left- or right-side walking when pedestrians move in opposite directions. Evaluating this effect under higher-density and crowded conditions would provide a further test of its applicability to realistic public-space settings.

\section*{CRediT authorship contribution statement}
YM: Conceptualization, Data curation, Formal analysis, Funding acquisition, Investigation, Supervision, Visualization, Writing – original draft, Writing – review \& editing.\\
TF: Investigation, Methodology, Resources, Writing – review \& editing.\\
YK: Investigation, Methodology, Writing – review \& editing.\\
TY: Investigation, Methodology, Writing – review \& editing.\\
MO: Investigation, Data curation, Writing – review \& editing.\\
KM: Investigation, Project administration, Writing – review \& editing.

\section*{Funding}
This work was supported by JST, ACT-X Grant Number JPMJAX21KH and PRESTO Grant Number JPMJPR25IA.

\section*{Declaration of competing interest}
The authors declare that they have no known competing financial interests or personal relationships that could have appeared to influence the work reported in this paper.

\section*{Data availability}
The data that support the findings of this study are not publicly available due to privacy and ethical restrictions.

\section*{Declaration of generative AI use}
During the preparation of this work, the authors used ChatGPT  (OpenAI) to assist with English-language editing and improve the readability of the manuscript. After using this tool, the authors reviewed and edited the content as needed and take full responsibility for the content of the published article.

\section*{Acknowledgements}
We sincerely thank the National Museum of Emerging Science and Innovation (Miraikan) for providing the experimental venues and for their generous organizational and operational support throughout the study.








\bibliographystyle{elsarticle-num}
\bibliography{citation}

@String{Computing = "Computing" }

@String{Computer = "{IEEE} Computer" }

@String{Springer = "Springer-Verlag" }

@article{allison1999effect,
  title={Effect of field size, head motion, and rotational velocity on roll vection and illusory self-tilt in a tumbling room},
  author={Allison, Robert S and Howard, Ian P and Zacher, James E},
  journal={Perception},
  volume={28},
  number={3},
  pages={299--306},
  year={1999},
  publisher={SAGE Publications Sage UK: London, England}
}

@article{palmisano1998stimulus,
  title={Stimulus eccentricity and spatial frequency interact to determine circular vection},
  author={Palmisano, Stephen and Gillam, Barbara},
  journal={Perception},
  volume={27},
  number={9},
  pages={1067--1077},
  year={1998},
  publisher={SAGE Publications Sage UK: London, England}
}

@article{guo2021effects,
  title={Effects of luminance contrast, averaged luminance and spatial frequency on vection},
  author={Guo, Xuanru and Nakamura, Shinji and Fujii, Yoshitaka and Seno, Takeharu and Palmisano, Stephen},
  journal={Experimental Brain Research},
  volume={239},
  number={12},
  pages={3507--3525},
  year={2021},
  publisher={Springer}
}

@article{nakamura1998stimulus,
  title={Stimulus size and eccentricity in visually induced perception of horizontally translational self-motion},
  author={Nakamura, Shinji and Shimojo, Shinsuke},
  journal={Perceptual and motor skills},
  volume={87},
  number={2},
  pages={659--663},
  year={1998},
  publisher={Sage Publications Sage CA: Los Angeles, CA}
}

@article{palmisano2015future,
  title={Future challenges for vection research: definitions, functional significance, measures, and neural bases},
  author={Palmisano, Stephen and Allison, Robert S and Schira, Mark M and Barry, Robert J},
  journal={Frontiers in psychology},
  volume={6},
  pages={193},
  year={2015},
  publisher={Frontiers Media SA}
}

@article{brandt1973differential,
  title={Differential effects of central versus peripheral vision on egocentric and exocentric motion perception},
  author={Brandt, Th and Dichgans, Johannes and Koenig, Ellen},
  journal={Experimental brain research},
  volume={16},
  number={5},
  pages={476--491},
  year={1973},
  publisher={Springer}
}

@article{tamada2015roles,
  title={Roles of size, position, and speed of stimulus in vection with stimuli projected on a ground surface},
  author={Tamada, Yasuaki and Seno, Takeharu},
  journal={Aerospace medicine and human performance},
  volume={86},
  number={9},
  pages={794--802},
  year={2015},
  publisher={Aerospace Medical Association}
}

@article{andersen1985induced,
  title={Induced self-motion in central vision.},
  author={Andersen, George J and Braunstein, Myron L},
  journal={Journal of Experimental Psychology: Human Perception and Performance},
  volume={11},
  number={2},
  pages={122},
  year={1985},
  publisher={American Psychological Association}
}

@inproceedings{furukawa2011vection,
  title={"Vection field" for pedestrian traffic control},
  author={Furukawa, Masahiro and Yoshikawa, Hiromi and Hachisu, Taku and Fukushima, Shogo and Kajimoto, Hiroyuki},
  booktitle={Proceedings of the 2nd augmented human international conference},
  pages={1--8},
  year={2011}
}

@inproceedings{ishii2016optical,
  title={Optical marionette: Graphical manipulation of human's walking direction},
  author={Ishii, Akira and Suzuki, Ippei and Sakamoto, Shinji and Kanai, Keita and Takazawa, Kazuki and Doi, Hiraku and Ochiai, Yoichi},
  booktitle={Proceedings of the 29th Annual Symposium on User Interface Software and Technology},
  pages={705--716},
  year={2016}
}

@inproceedings{sakamoto2019guided,
  title={Guided walking to direct pedestrians toward the same destination},
  author={Sakamoto, Nobuhito and Furukawa, Masahiro and Kurokawa, Masataka and Maeda, Taro},
  booktitle={Proceedings of the 10th Augmented Human International Conference 2019},
  pages={1--8},
  year={2019}
}

@misc{mikawa-APMAR2025,
  title={Realizing Large Spatial Display in Public Space using Multiple Discrete Signages},
  author={Mikawa, Yuri and Kubota, Yuki and Fukiage, Taiki and Yokosaka, Takumi and Ogawa, Maki and Maruya, Kazushi},
  howpublished={Pitch Your Work in the 17th Asia-Pacific Workshop on Mixed and Augmented Reality (APMAR 2025)},
  year={2025},
  month={September},
  note={}
}

@incollection{shimizu2024gentlepoles,
  title={GentlePoles: Designing Wooden Pole Actuators for Guiding People},
  author={Shimizu, Masaya and Linde, Berend te and Yoshida, Takatoshi and Horie, Arata and Hanamitsu, Nobuhisa and Minamizawa, Kouta},
  booktitle={SIGGRAPH Asia 2024 Posters},
  pages={1--2},
  year={2024}
}

@mastersthesis{shimizu2024master,
  author = {Shimizu, Masaya},
  title  = {GentlePoles: design of rotating wooden pole actuators for ambient and gentle guidance},
  school = {Graduate School of Media Design, Keio University},
  year   = {2024},
  type   = {Master's Thesis},
}

@inproceedings{gomi2019innovative,
  title={Innovative mobile force display: Buru-Navi},
  author={Gomi, Hiroaki and Ito, Sho and Tanase, Ryoma},
  booktitle={26th International Display Workshops, IDW 2019},
  pages={962--965},
  year={2019},
  organization={International Display Workshops}
}

@inproceedings{kataoka2016dynamic,
  title={Dynamic guide signs system to control pedestrian flow},
  author={Kataoka, Haruno and Hashiguchi, Kyoko and Wago, Kae and Ichikawa, Yusuke and Tezuka, Hirohisa and Yamashita, Shinichiro and Kuhara, Yusaku and Akiyama, Tetsuo},
  booktitle={Proceedings of the 2016 ACM international joint conference on pervasive and ubiquitous computing: Adjunct},
  pages={1572--1577},
  year={2016}
}

@article{arikawa2007navitime,
  title={Navitime: Supporting pedestrian navigation in the real world},
  author={Arikawa, Masatoshi and Konomi, Shin'ichi and Ohnishi, Keisuke},
  journal={IEEE Pervasive Computing},
  volume={6},
  number={3},
  pages={21--29},
  year={2007},
  publisher={IEEE}
}

@inproceedings{suzuki2019midair,
  title={Midair hand guidance by an ultrasound virtual handrail},
  author={Suzuki, Shun and Fujiwara, Masahiro and Makino, Yasutoshi and Shinoda, Hiroyuki},
  booktitle={2019 IEEE World Haptics Conference (WHC)},
  pages={271--276},
  year={2019},
  organization={IEEE}
}

@inproceedings{tsukada2004activebelt,
  title={Activebelt: Belt-type wearable tactile display for directional navigation},
  author={Tsukada, Koji and Yasumura, Michiaki},
  booktitle={international conference on ubiquitous computing},
  pages={384--399},
  year={2004},
  organization={Springer}
}

@inproceedings{strachan2005gpstunes,
  title={GpsTunes: controlling navigation via audio feedback},
  author={Strachan, Steven and Eslambolchilar, Parisa and Murray-Smith, Roderick and Hughes, Stephen and O'Modhrain, Sile},
  booktitle={Proceedings of the 7th international conference on Human computer interaction with mobile devices \& services},
  pages={275--278},
  year={2005}
}

@inproceedings{gleeson2009communication,
  title={Communication of direction through lateral skin stretch at the fingertip},
  author={Gleeson, Brian T and Horschel, Scott K and Provancher, William R},
  booktitle={World Haptics 2009-Third Joint EuroHaptics conference and Symposium on Haptic Interfaces for Virtual Environment and Teleoperator Systems},
  pages={172--177},
  year={2009},
  organization={IEEE}
}

@article{sousa2009head,
  title={Head-mounted display versus desktop for 3D navigation in virtual reality: a user study},
  author={Sousa Santos, Beatriz and Dias, Paulo and Pimentel, Angela and Baggerman, Jan-Willem and Ferreira, Carlos and Silva, Samuel and Madeira, Joaquim},
  journal={Multimedia tools and applications},
  volume={41},
  number={1},
  pages={161--181},
  year={2009},
  publisher={Springer}
}

@article{hanssens2013visually,
  title={Visually induced postural reactivity is velocity-dependent at low temporal frequencies and frequency-dependent at high temporal frequencies},
  author={Hanssens, J-M and Allard, R and Giraudet, G and Faubert, J},
  journal={Experimental brain research},
  volume={229},
  number={1},
  pages={75--84},
  year={2013},
  publisher={Springer}
}

@inproceedings{sugiura2017experimental,
  title={Experimental study on control of visually evoked postural responses by galvanic vestibular stimulation},
  author={Sugiura, Akihiro and Akachi, Kazuya and Ito, Akira Yoshida Chiharu and Kondo, Shuya and Tanaka, Kunihiko and Takada, Hiroki},
  booktitle={2017 12th International Conference on Computer Science and Education (ICCSE)},
  pages={77--82},
  year={2017},
  organization={IEEE}
}

@article{guerraz2008mechanisms,
  title={Mechanisms underlying visually induced body sway},
  author={Guerraz, Michel and Bronstein, Adolfo M},
  journal={Neuroscience letters},
  volume={443},
  number={1},
  pages={12--16},
  year={2008},
  publisher={Elsevier}
}

@article{mills2007dynamic-hugarian,
  title={The dynamic hungarian algorithm for the assignment problem with changing costs},
  author={Mills-Tettey, G Ayorkor and Stentz, Anthony and Dias, M Bernardine},
  journal={Robotics Institute, Pittsburgh, PA, Tech. Rep. CMU-RI-TR-07-27},
  volume={7},
  year={2007}
}

@inproceedings{mirgorodskaya2020using,
  title={Using digital signage technologies in retail marketing activities},
  author={Mirgorodskaya, Olga and Ivanchenko, Olesya and Dadayan, Narine},
  booktitle={Proceedings of the International Scientific Conference-Digital Transformation on Manufacturing, Infrastructure and Service},
  pages={1--7},
  year={2020}
}

@article{sauvan1995spatiotemporal,
  title={Spatiotemporal boundaries of linear vection},
  author={Sauvan, Xavier M and Bonnet, Claude},
  journal={Perception \& Psychophysics},
  volume={57},
  number={6},
  pages={898--904},
  year={1995},
  publisher={Springer}
}

@inproceedings{yu2010perception,
  title={Perception of linear self motion under different visual contrast conditions for development of driving simulator},
  author={Yu, Yinghua and Wu, Jinglong},
  booktitle={IEEE/ICME International Conference on Complex Medical Engineering},
  pages={115--118},
  year={2010},
  organization={IEEE}
}

@article{sauvan1993properties,
  title={Properties of curvilinear vection},
  author={Sauvan, Xavier M and Bonnet, Claude},
  journal={Perception \& Psychophysics},
  volume={53},
  number={4},
  pages={429--435},
  year={1993},
  publisher={Springer}
}

@article{riecke2011compelling,
  title={Compelling self-motion through virtual environments without actual self-motion: using self-motion illusions (“vection”) to improve user experience in VR},
  author={Riecke, Bernhard E},
  journal={Virtual reality},
  volume={8},
  number={1},
  pages={149--178},
  year={2011},
  publisher={InTech Rijeka, Croatia}
}

@article{wang2021visually,
  title={Visually induced roll circular vection: Do effects of stimulation velocity differ for supine and upright participants?},
  author={Wang, Yixuan and Du, Bo and Wei, Yue and So, Richard HY},
  journal={Frontiers in Virtual Reality},
  volume={2},
  pages={611214},
  year={2021},
  publisher={Frontiers Media SA}
}

@article{seno2009,
  title={The introduction and application to VR contents of the stimulus parameters that effectively induce vection and Brain imaging studies of vection},
  author={Takeharu Seno},
  journal={Transactions of the Virtual Reality Society of Japan},
  volume={14},
  number={4},
  pages={481--490},
  year={2009},
  publisher={The Virtual Reality Society of Japan}
}

@article{palmisano2004jitter,
  title={Jitter and size effects on vection are immune to experimental instructions and demands},
  author={Palmisano, Stephen and Chan, Amy YC},
  journal={Perception},
  volume={33},
  number={8},
  pages={987--1000},
  year={2004},
  publisher={SAGE Publications Sage UK: London, England}
}

@article{palmisano2012horizontal,
  title={Horizontal fixation point oscillation and simulated viewpoint oscillation both increase vection in depth},
  author={Palmisano, Stephen and Kim, Juno and Freeman, Tom CA},
  journal={Journal of Vision},
  volume={12},
  number={12},
  pages={15--15},
  year={2012},
  publisher={The Association for Research in Vision and Ophthalmology}
}

@article{seno2017oscillating,
  title={The oscillating potential model of visually induced vection},
  author={Seno, Takeharu and Sawai, Ken-ichi and Kanaya, Hidetoshi and Wakebe, Toshihiro and Ogawa, Masaki and Fujii, Yoshitaka and Palmisano, Stephen},
  journal={i-Perception},
  volume={8},
  number={6},
  pages={2041669517742176},
  year={2017},
  publisher={Sage Publications Sage UK: London, England}
}

@article{palmisano2011simulated,
  title={Simulated viewpoint jitter shakes sensory conflict accounts of vection},
  author={Palmisano, Stephen and Kim, Juno and Allison, Robert and Bonato, Frederick},
  journal={Seeing and perceiving},
  volume={24},
  number={2},
  pages={173--200},
  year={2011},
  publisher={Brill}
}

@article{kitazaki2003attentional,
  title={Attentional modulation of self-motion perception},
  author={Kitazaki, Michiteru and Sato, Takao},
  journal={Perception},
  volume={32},
  number={4},
  pages={475--484},
  year={2003},
  publisher={SAGE Publications Sage UK: London, England}
}

@article{nakamura2006effects,
  title={Effects of depth, eccentricity and size of additional static stimulus on visually induced self-motion perception},
  author={Nakamura, Shinji},
  journal={Vision Research},
  volume={46},
  number={15},
  pages={2344--2353},
  year={2006},
  publisher={Elsevier}
}

@article{andersen2000neural,
  title={Neural mechanisms for self-motion perception in area MST},
  author={Andersen, Richard A and Shenoy, Krishna V and Crowell, James A and Bradley, David C},
  journal={International Review of Neurobiology},
  volume={44},
  pages={219--233},
  year={2000},
  publisher={Elsevier}
}

@article{ash2011vection,
  title={Vection in depth during consistent and inconsistent multisensory stimulation},
  author={Ash, April and Palmisano, Stephen and Kim, Juno},
  journal={Perception},
  volume={40},
  number={2},
  pages={155--174},
  year={2011},
  publisher={SAGE Publications Sage UK: London, England}
}

@article{palmisano2018search,
  title={The search for instantaneous vection: An oscillating visual prime reduces vection onset latency},
  author={Palmisano, Stephen and Riecke, Bernhard E},
  journal={PloS one},
  volume={13},
  number={5},
  pages={e0195886},
  year={2018},
  publisher={Public Library of Science San Francisco, CA USA}
}

@article{kovacs2008neural,
  title={Neural correlates of visually induced self-motion illusion in depth},
  author={Kov{\'a}cs, Gyula and Raabe, Markus and Greenlee, Mark W},
  journal={Cerebral cortex},
  volume={18},
  number={8},
  pages={1779--1787},
  year={2008},
  publisher={Oxford University Press}
}

@incollection{dichgans1978visual,
  title={Visual-vestibular interaction: Effects on self-motion perception and postural control},
  author={Dichgans, Johannes and Brandt, Thomas},
  booktitle={Perception},
  pages={755--804},
  year={1978},
  publisher={Springer}
}

@incollection{deangelis2012visual,
  author    = {DeAngelis, Gregory C. and Angelaki, Dora E.},
  title     = {Visual--Vestibular Integration for Self-Motion Perception},
  booktitle = {The Neural Bases of Multisensory Processes},
  editor    = {Murray, Micah M. and Wallace, Mark T.},
  publisher = {CRC Press/Taylor \& Francis},
  address   = {Boca Raton, FL},
  year      = {2012},
  chapter   = {31},
  series    = {Frontiers in Neuroscience}
}

@article{chen2011convergence,
  title={Convergence of vestibular and visual self-motion signals in an area of the posterior sylvian fissure},
  author={Chen, Aihua and DeAngelis, Gregory C and Angelaki, Dora E},
  journal={Journal of Neuroscience},
  volume={31},
  number={32},
  pages={11617--11627},
  year={2011},
  publisher={Society for Neuroscience}
}

@article{thurrell2002vection,
  title={Vection increases the magnitude and accuracy of visually evoked postural responses},
  author={Thurrell, A. E. I. and Bronstein, A. M.},
  journal={Experimental Brain Research},
  volume={147},
  number={4},
  pages={558--561},
  year={2002},
  doi={10.1007/s00221-002-1266-y}
}

@article{kawakita2000body,
  title={Body sway induced by depth linear vection in reference to central and peripheral visual field},
  author={Kawakita, T. and Kuno, S. and Ninomija, H. and Miyao, M.},
  journal={Japanese Journal of Physiology},
  volume={50},
  number={3},
  pages={315--321},
  year={2000},
  doi={10.2170/jjphysiol.50.315}
}

@article{kunihiro2004postural,
  title={Postural sway during optokinetic stimulation: The relationship between vection and posture},
  author={Kunihiro, T. and others},
  journal={Acta Oto-Laryngologica},
  volume={124},
  number={6},
  pages={637--643},
  year={2004},
  doi={10.1080/00016480410016910}
}

@inproceedings{khan2014dbscan,
  title={DBSCAN: Past, present and future},
  author={Khan, Kamran and Rehman, Saif Ur and Aziz, Kamran and Fong, Simon and Sarasvady, Sababady},
  booktitle={The fifth international conference on the applications of digital information and web technologies (ICADIWT 2014)},
  pages={232--238},
  year={2014},
  organization={IEEE}
}

@article{galea2017international,
  title={An international survey and full-scale evacuation trial demonstrating the effectiveness of the active dynamic signage system concept},
  author={Galea, ER and Xie, H and Deere, S and Cooney, D and Filippidis, L},
  journal={Fire and Materials},
  volume={41},
  number={5},
  pages={493--513},
  year={2017},
  publisher={Wiley Online Library}
}

@article{galea2017evaluating,
  title={Evaluating the effectiveness of an improved active dynamic signage system using full scale evacuation trials},
  author={Galea, Edwin R and Xie, Hui and Deere, Steven and Cooney, David and Filippidis, Lazaros},
  journal={Fire Safety Journal},
  volume={91},
  pages={908--917},
  year={2017},
  publisher={Elsevier}
}

@article{jamshidi2020wayfinding,
  title={Wayfinding in interior environments: An integrative review},
  author={Jamshidi, Saman and Ensafi, Mahnaz and Pati, Debajyoti},
  journal={Frontiers in Psychology},
  volume={11},
  pages={549628},
  year={2020},
  publisher={Frontiers Media SA}
}

@article{mishler2017improving,
  title={Improving wayfinding for older users with selective attention deficits},
  author={Mishler, Ada D and Neider, Mark B},
  journal={ergonomics in design},
  volume={25},
  number={1},
  pages={11--16},
  year={2017},
  publisher={SAGE Publications Sage CA: Los Angeles, CA}
}

@article{galea2014experimental,
  title={Experimental and survey studies on the effectiveness of dynamic signage systems},
  author={Galea, Edwin R and Xie, Hui and Lawrence, Peter J},
  journal={Fire Safety Science},
  volume={11},
  pages={1129--1143},
  year={2014},
  publisher={International Association for Fire Safety Science}
}

@article{fuller2002arrow,
  title={The arrow--directional semiotics: Wayfinding in transit},
  author={Fuller, Gillian},
  journal={Social semiotics},
  volume={12},
  number={3},
  pages={231--244},
  year={2002},
  publisher={Taylor \& Francis}
}

@article{zwahlen1999legibility,
  title={Legibility of traffic sign text and symbols},
  author={Zwahlen, Helmut T and Schnell, Thomas},
  journal={Transportation Research Record},
  volume={1692},
  number={1},
  pages={142--151},
  year={1999},
  publisher={SAGE Publications Sage CA: Los Angeles, CA}
}

@article{helbing2001self,
  title={Self-organizing pedestrian movement},
  author={Helbing, Dirk and Moln{\'a}r, P{\'e}ter and Farkas, Ill{\'e}s J and Bolay, Kai},
  journal={Environment and planning B: planning and design},
  volume={28},
  number={3},
  pages={361--383},
  year={2001},
  publisher={SAGE Publications Sage UK: London, England}
}

@article{zanlungo2012microscopic,
  title={A microscopic “social norm” model to obtain realistic macroscopic velocity and density pedestrian distributions},
  author={Zanlungo, Francesco and Ikeda, Tetsushi and Kanda, Takayuki},
  journal={PloS one},
  volume={7},
  number={12},
  pages={e50720},
  year={2012},
  publisher={Public Library of Science San Francisco, USA}
}

@article{murtagh2021outdoor,
  author  = {Murtagh, Elaine M. and Mair, Jacqueline L. and
             Aguiar, Elroy and Tudor-Locke, Catrine and
             Murphy, Marie H.},
  title   = {Outdoor Walking Speeds of Apparently Healthy Adults:
             A Systematic Review and Meta-analysis},
  journal = {Sports Medicine},
  year    = {2021},
  volume  = {51},
  pages   = {125--141},
  doi     = {10.1007/s40279-020-01351-3}
}





\end{document}